\documentclass{article}

\pdfoutput=1
\usepackage{arxiv}
\usepackage{color}
\usepackage[utf8]{inputenc}
\usepackage{url}
\usepackage{hyperref}
\usepackage{listings}
\usepackage{siunitx}
\usepackage{amsmath}
\usepackage{cleveref}
\usepackage{nicefrac}
\usepackage{graphicx}
\usepackage[square,sort,comma,numbers]{natbib}

\usepackage{epsfig}
\usepackage{float}
\usepackage[section]{placeins}

\usepackage[labelformat=empty]{caption}

\title{Awareness as inference in a higher-order state space}
 \author{Stephen M. Fleming  \\
 Wellcome Centre for Human Neuroimaging \\
 12 Queen Square \\
 London WC1N 3AR \\
 University College London \\
 \texttt{stephen.fleming@ucl.ac.uk} \\
 }
 \date{\today}

\begin{document}
 \maketitle

\begin{abstract}
Humans have the ability to report the contents of their subjective experience -- we can say to each other, ``I am aware of \emph{X}". The decision processes that support these reports about mental contents remain poorly understood. In this article I propose a computational framework that characterises awareness reports as metacognitive decisions (inference) about a generative model of perceptual content. This account is motivated from the perspective of how flexible hierarchical state spaces are built during learning and decision-making. Internal states supporting awareness reports, unlike those covarying with perceptual contents, are simple and abstract, varying along a one-dimensional continuum from absent to present. A critical feature of this architecture is that it is both higher-order and asymmetric: a vast number of perceptual states is nested under ``present", but a much smaller number of possible states nested under ``absent". Via simulations I show that this asymmetry provides a natural account of observations of ``global ignition" in brain imaging studies of awareness reports.
\end{abstract}

\newcommand{\ie}{i.e.~\/}
\newcommand{\eg}{e.g.~\/}
\newcommand{\rleft}{\raggedleft}
\newcommand{\tn}{\tabularnewline}

\setcounter{page}{1}

\section{Introduction}

Humans have the ability to report the contents of their subjective experience - we can say to each other, ``I am aware of \emph{X}". Such reports, unlike many other aspects of behaviour, are intended to convey information about experience \cite{Frith:1999}. This property of awareness reports makes them central to a science of consciousness, which has focused on measuring and quantifying differences in awareness while holding other aspects of stimuli and behavioural performance constant \cite{Baars:1993, Dehaene:2011}. In this article I propose a computational framework that characterises awareness reports as metacognitive decisions (inference) about a generative model of perceptual content \cite{Fleming:2017}. This higher-order state space (HOSS) framework builds on Bayesian approaches to perception that invoke hierarchical probabilistic inference as a route towards efficiently modeling the external world \cite{Friston:2005, Helmholtz:1860, Hohwy:2013, Kersten:2004}.

The outline of the paper is as follows. I start by describing the psychological processes hypothesised to support awareness reports with reference to experimental paradigms commonly used to study conscious awareness. Second, I outline the central hypothesis, that awareness is a higher-order state in a generative model of perceptual contents. Third, I model a simple perceptual decision to explicate aspects of the framework, and distinguish it from other, related approaches such as signal detection theory (SDT; \cite{Green:1966, King:2014}). Finally, I highlight empirical predictions that flow from the model, and how it relates to existing theories of consciousness such as global workspace and higher-order theories.

\section{Psychological basis of awareness reports}

Several authors have proposed that the psychological basis of a (visual) awareness report is an internal decision about the \emph{visibility} of perceptual contents\footnote{The same computational considerations likely hold for awareness of other sensory modalities -- a focus on visibility here reflects a historical bias towards vision in studies of conscious awareness.} \cite{Sergent:2004, Ramsoy:2004, King:2014}. This implies that internal states supporting awareness reports, unlike those covarying with perceptual contents themselves, are both \emph{simple} and \emph{abstract} -- simple because they vary along a one-dimensional continuum from unaware to aware, and abstract because they do not encode the perceptual state itself, only its presence or absence. Note that a better terminology for ``unaware" is really ``absent", ``unseen" or ``noise", as participants remain aware of seeing nothing on trials on which they report ``unaware". Awareness reports also \emph{refer to} different subsets of perceptual content: for instance, subjects may be asked ``did you see the word?", ``did you see the number?" or ``did you see anything at all?". These two features imply that awareness reports are metacognitive decisions \emph{about} a rich perceptual generative model. I will make this hypothesis more concrete in the next section.

A range of experimental paradigms have been developed to introduce variability in awareness reports while keeping other aspects of stimuli and behaviour fixed (see \cite{Kim:2005} for a review). For example, using backward masking, Dehaene and colleagues found that they could make words invisible while showing (via priming effects and brain imaging) that they were processed up to a semantic level \cite{Dehaene:2001}. When subjects reported consciously seeing the words, whole-brain fMRI showed elevated activations in the parietal and prefrontal cortex, which have become known as ``global ignition" responses due to their non-linear response profile in relation to stimulation strength \cite{Dehaene:2011, Del-Cul:2007}.\footnote{Since these classic studies, alternative explanations of frontoparietal ignition have been put forward, including that it is involved in the act of reporting, but not conscious awareness, or that it reflects greater performance capacity on conscious trials \cite{Aru:2012, Lau:2006}. These debates are ongoing (see \cite{Tsuchiya:2015, Michel:}, for recent arguments from both sides).}

\section{Hypothesis}

\begin{figure}
  \centering
  \includegraphics[width=160mm]{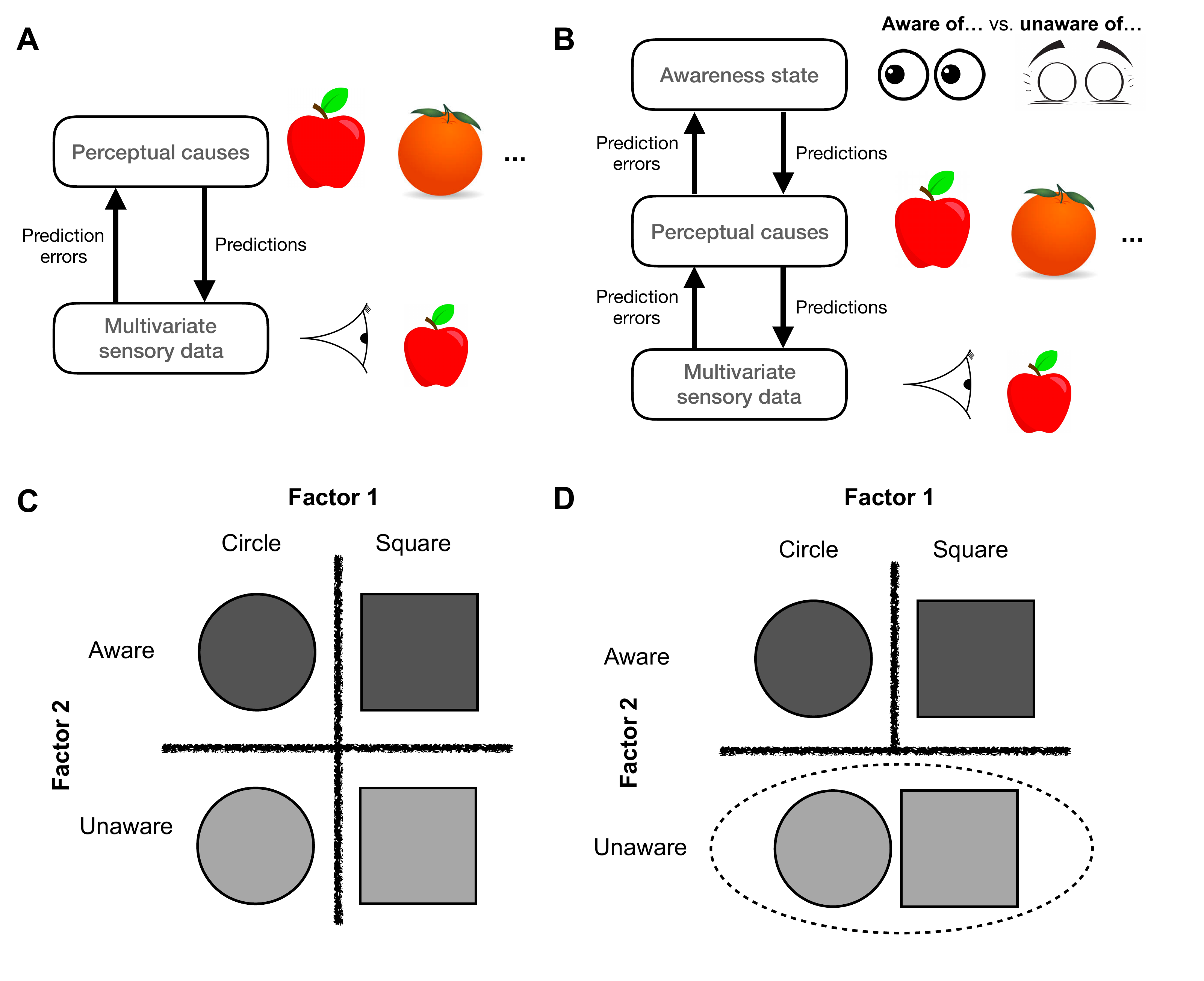}
  \caption{Figure 1. A) Schematic of standard predictive coding archictures in which hypotheses about perceptual causes (predictions) are updated in response to prediction errors generated by incoming sensory data. B) Extension of the predictive coding architecture in (A) to accommodate a higher-order awareness state. C) Factorisation of awareness and perceptual content. D) Illustration of the asymmetry that ensues from factorising an awareness state; unawareness of a circle is a similar state to unawareness of a square.}
  \label{fig:fig1}
\end{figure}

In common with other predictive processing approaches, we assume that the brain is engaged in building a hierarchical, probabilistic generative model of the world, one in which inference and learning proceed using (approximations of) Bayes' rule. One algorithmic implementation of Bayesian generative models is predictive coding, whereby perceptual causes encoded at higher levels of the system generate predictions about incoming sensory data \cite{Friston:2005, Hohwy:2013}. Figure \ref{fig:fig1}A presents an outline of this scheme, in which two perceptual causes (apple or orange) generate predictions that ``compete" to explain the incoming sensory data, while lower layers in turn signal the error in the current prediction. Via a prediction error minimization scheme, over time the best explanation of the sensory data is ``selected" at higher levels of the system.

The novel aspect of the current framework is its focus on incorporating \emph{awareness} into the perceptual generative model -- explaining how decisions to respond ``I am aware of \emph{X}" or ``I am unaware of \emph{X}" get made.\footnote{Note that here I am focusing on \emph{reportable} states of awareness, and leaving aside the issue of whether non-reportable contents may be conscious \cite{Block:1995, Block:2011}. By adopting this stance, we can frame a clear question that is answerable by cognitive science: what are the computational processes involved in making awareness reports? \cite{Dennett:1993, Graziano:2013}}

The central hypothesis is:

\begin{quotation}

Awareness is a higher-order state in a generative model of perceptual contents.

\end{quotation}

Awareness reports are governed by a second-order (metacognitive) inference about the state of a first-order (perceptual) generative model \cite{Fleming:2017}. One way of implementing this second-order inference is by adding an additional hierarchical state above the perceptual generative model, which I refer to as an ``awareness state" (Figure \ref{fig:fig1}B). Paralleling the psychological simplicity of awareness reports, the awareness state is also simple, and signals a probability of whether there is perceptual content in the lower layers (corresponding to reports of ``present" or ``absent"). It is also part of the generative model, such that if the model is run forward, states of presence (vs. absence) lead to the top-down generation of perceptual states in lower layers.

As we reviewed above, a central property of awareness states is that they are abstract -- we can flexibly interrogate awareness of not only apples or oranges, but also many other aspects of perceptual experience. In the language of probabilistic generative models, this implies awareness states are factorised with respect to lower-order perceptual causes. This notion of factorisation is depicted in Figure \ref{fig:fig1}C for the case of awareness of different shapes (a circle or square). Rather than maintaining separate states for aware-of-circle, aware-of-square, unaware-of-circle and unaware-of-square, this space can be factorised into two states, one for circle/square, and another for aware/unaware. However, this factorisation is \emph{asymmetric} (Figure \ref{fig:fig1}D). In the absence of awareness, there is (by definition) an absence of perceptual content, such that being unaware of a circle is similar to being unaware of a square. In contrast, a large number of potential perceptual states is nested under the awareness state of ``presence". This imposes an asymmetry in the model architecture which we will leverage in the next section when seeking to account for global ignition responses.

\section{Model}

\begin{figure}
  \centering
  \includegraphics[width=160mm]{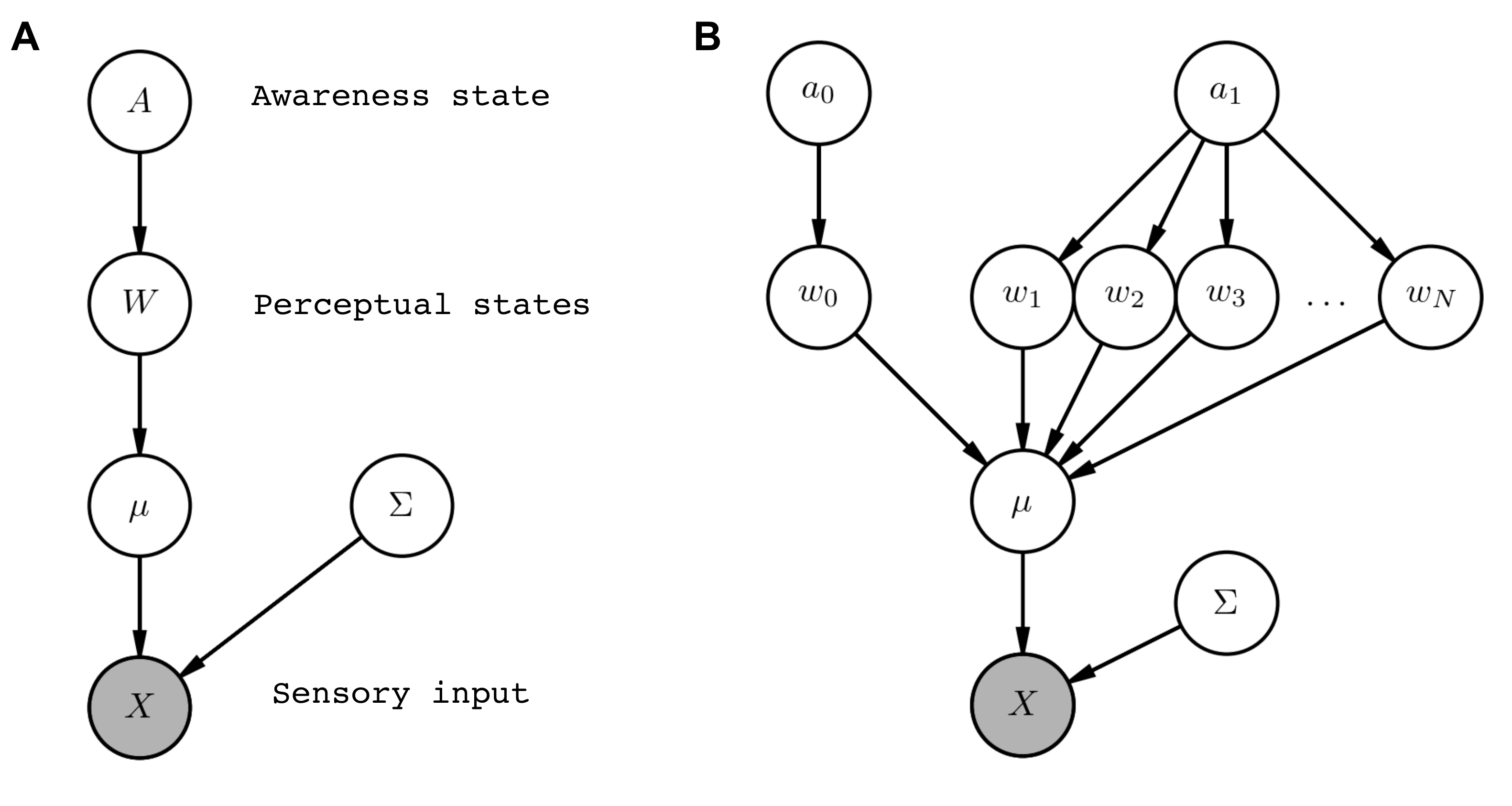}
  \caption{Figure 2. A) Probabilistic graphical model of awareness reports. Nodes represent random variables and the graph structure is used to indicate dependencies as indicated by directed arrows. The shaded node indicates that this variable is observed by the system as sensory input. B) Expanded version of the graphical model from panel (A) that makes explicit the asymmetry in the state space. Figures created using the Daft package in Python.}
  \label{fig:fig2}
\end{figure}

The model can be described formally in terms of a probabilistic graphical model \cite{Pearl:1988}, where nodes correspond to unknown variables and the graph structure is used to indicate dependencies between variables. These graphs provide a concise description of how sensory data is generated (Figure \ref{fig:fig2}A).

$W$ is a $1 \times N$ vector that encodes the relative probabilities of each of $N$ discrete perceptual states. $A$ is a scalar awareness state encoding the probability of a perceptual state $W$ being ``present" ($w_1 \dots w_N$) or ``absent" ($w_0$). Each ``perceptual" state $W$ here is discrete, but in reality this state space is likely to be multidimensional and also hierarchically organised. By expanding the graphical model to ennumerate each discrete state (Figure \ref{fig:fig2}B), it is straightforward to see how this architecture imposes an asymmetry on the perceptual state space: awareness ($a_1$) entails the possibility of perceptual content ($w_1 \dots w_N$), whereas unawareness leads to the absence of content ($w_0$). To simulate multivariate sensory data ($X$), $W$ in turn determines the value of $\mu$, which is a $M \times N$ matrix defining the location (mean) of a multivariate Gaussian in a feature space of dimensionality $M$. $\Sigma$ is a $M \times M$ covariance matrix which in the current simulations is fixed and independent of $A$ and $W$.

When answering the query, ``Present or absent?", we compute the posterior $P(A|X=x)$, marginalising over perceptual states $W$:

\begin{equation}
\begin{split}
P(A|X=x) & \propto \displaystyle\sum_{W}  P(A)P(W|A)P(X=x|W) \\
& \propto  P(A) \displaystyle\sum_{W}  P(W|A)P(X=x|W) \\
\end{split}
\end{equation}

where the likelihood of $X$ given $W$ is:

\begin{equation}
P(X=x|W) \sim N(\mu_W, \Sigma)
\end{equation}

As in standard models of perceptual decision-making such as SDT, inference on contents $W$ is also straightforward:

\begin{equation}
P(W|X=x) \propto \displaystyle\sum_{A}  P(A)P(W|A)P(X=x|W)
\end{equation}

\subsection{Simulations}

\begin{figure}
  \centering
  \includegraphics[width=160mm]{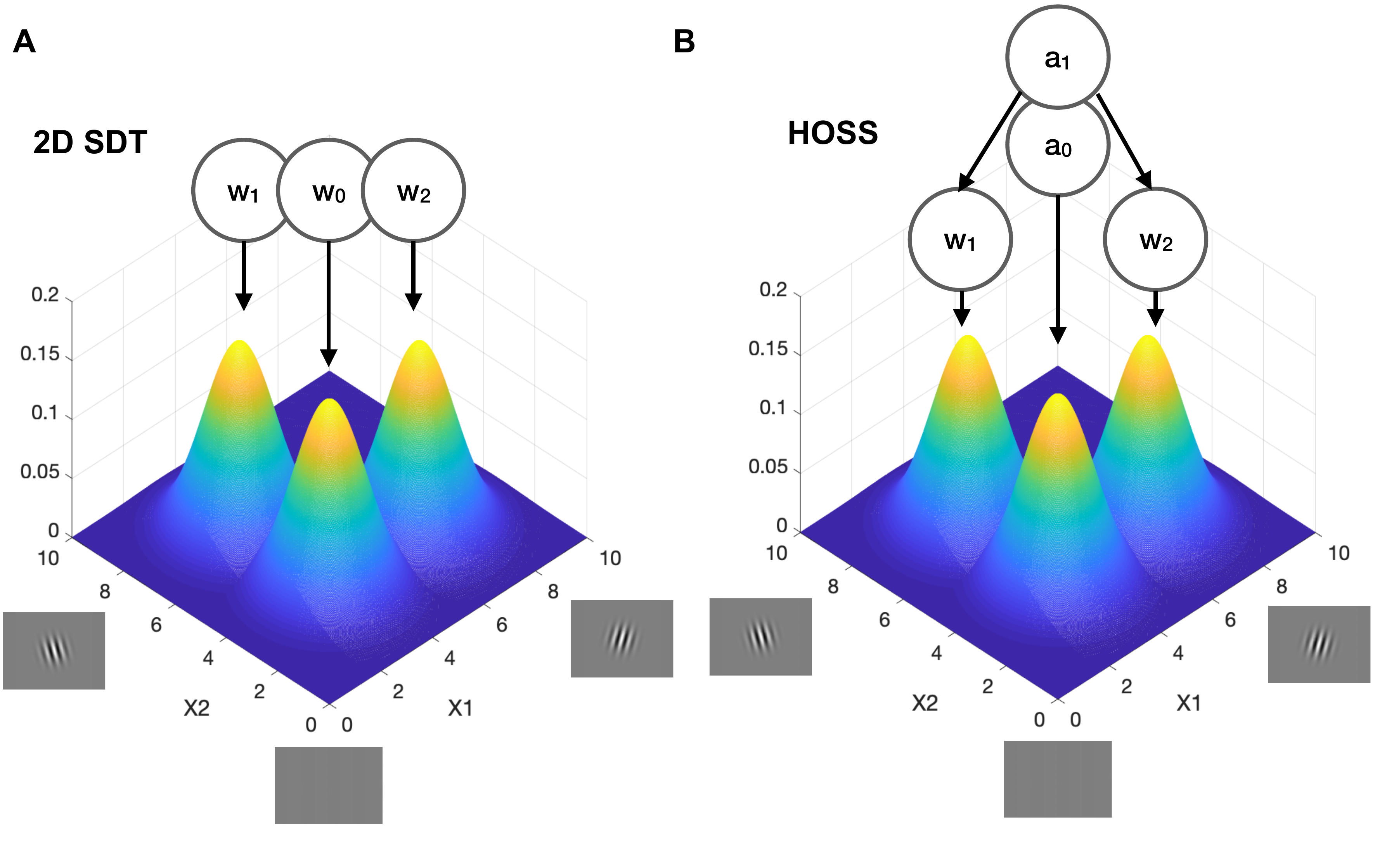}
  \caption{Figure 3. A) Two-dimensional feature space for a toy perceptual decision problem involving classifying two possible stimuli (e.g. left- and right-tilted Gabors). Each Gaussian indicates the likelihood of observing a pair of features (e.g. orientation) given each stimulus class (where $\mu = [3.5\;7]$, $[7\;3.5]$ or $[3.5\;3.5]$ and $\Sigma$ is the identity matrix). The right-tilted stimuli occupy the righthand side of the grid; left-tilted stimuli occupy lefthand side of the grid. The absence of stimulation is represented by a distribution in which activation of each feature is low, towards the origin. In two-dimensional signal detection theory (SDT), there are three stimulus classes organised in a flat (non-hierarchical) structure. B) The same two-dimensional feature space from (A), modified to make explicit the hierarchical aspect of the higher-order state space (HOSS) model. A higher-order awareness state ($a_1$) nests perceptual states $w_1$ and $w_2$.}
  \label{fig:fig3}
\end{figure}

To simulate the model I build on previous work using a two-dimensional feature space to capture important features of multidimensional perceptual categorisation \cite{King:2014}. Each axis represents the strength of activation of one of two possible stimulus features, such as leftward or rightward tilted grating orientations (see Figure \ref{fig:fig3}). The origin represents low activation on both features, consistent with no stimulus (or noise) being presented. As in the more general case described in the previous section, each stimulus category generates samples from a multivariate Gaussian whose mean is dominated by one or other feature. Thus if I receive a sample of $X = [2\;0]$, I can be confident that I was shown a left-tilted stimulus; if I receive a sample $X = [0\;2]$, I can be confident in seeing a right-tilted stimulus.

King and Dehaene \cite{King:2014} showed that by placing different types of decision criteria onto this space, multiple empirical relationships between discrimination performance, confidence and visibility could be simulated. In their model, visibility was modeled as the distance from the origin, and stimulus awareness reflected a first-order (flat) perceptual categorisation in which ``absent" was one of several potential stimulus classes (Figure \ref{fig:fig3}A). Our model builds closely on theirs and inherits the benefits of being able to accommodate dissociations between forced-choice responding and awareness reports. However it differs in proposing that awareness is not inherent to perceptual categorisation; instead, perceptual categorisation is \emph{nested} under a generative model of awareness (Figure \ref{fig:fig3}B). In other words, unlike in SDT, deciding that a stimulus is ``absent" in the HOSS model is governed by a more abstract state than deciding a stimulus is tilted to the left or right. We will see that this seemingly minor change in architecture leads to important consequences for the relationship between awareness and global ignition.

\begin{figure}
  \centering
  \includegraphics[width=160mm]{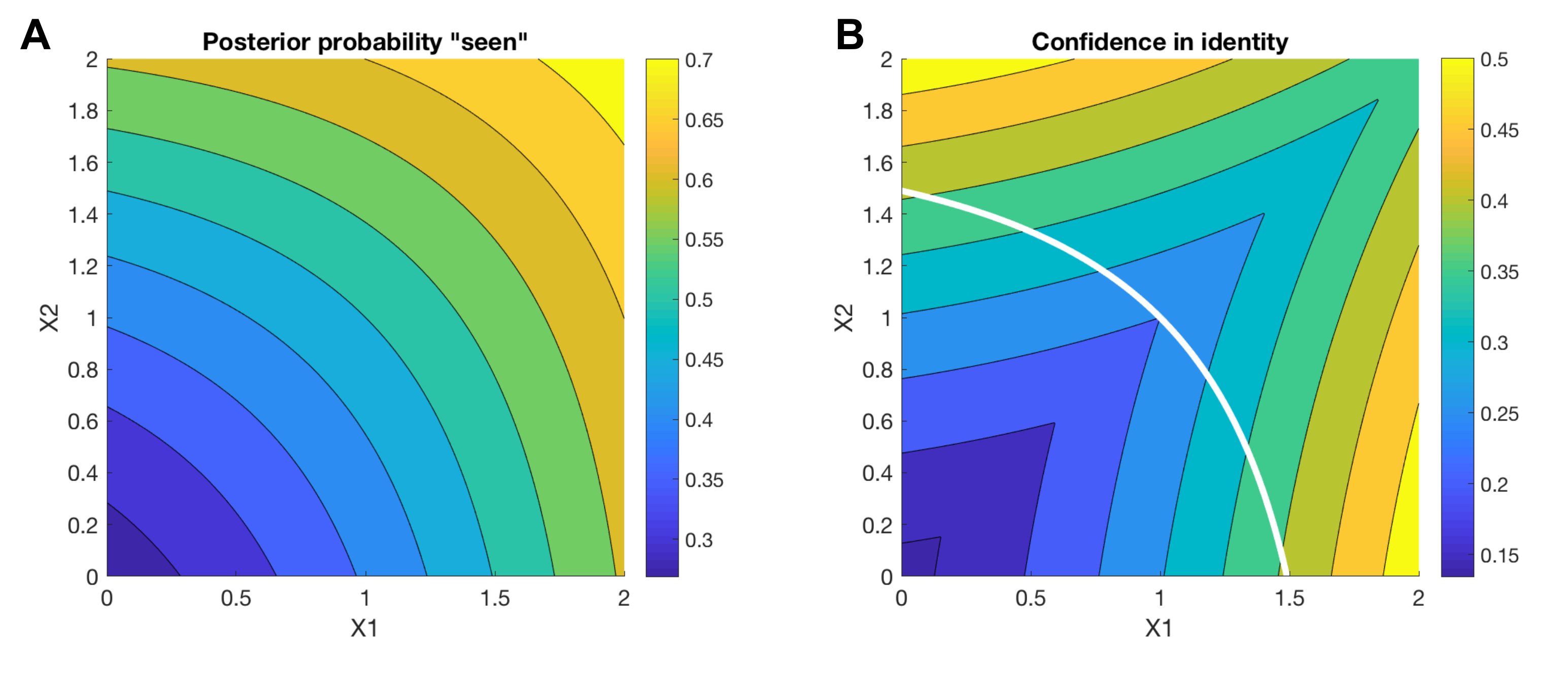}
  \caption{Figure 4. Simulations of inference on A) awareness state $A$ and B) perceptual states $W$, as a function of sensory input $X$ (where $\mu = [0.5\;1.5]$, $[1.5\;0.5]$ or $[0.5\;0.5]$ and $\Sigma$ is the identity matrix). In panel (A) the posterior probability of a report of ``presence" rises from the lower left to the upper right of the grid. In panel (B) confidence in stimulus identify (e.g. left- or right-tilted Gabor) increases towards the corners of the grid. Overlaid in white is the 0.5 contour from panel (A) showing that graded changes in confidence in identity still occur on trials that have a high likelihood of being classed as ``unseen" by the model. Confidence in identity was computed as $\max[P(w_1|X=x), P(w_2|X=x)]$.}
  \label{fig:fig4}
\end{figure}

To explore the properties of the model I simulate inference at different levels of the hierarchy for the two-class stimulus discrimination problem described in Figure \ref{fig:fig3}B. I first simulate, for a variety of two-dimensional inputs ($X$'s), the probability of saying ``aware" or ``seen" ($P(A=a_1|X=x)$). Figure \ref{fig:fig4}A shows that this probability rises in a graded manner from the lower left corner of the graph (low activation of any feature) to the upper right (high activation of both features). In contrast, confidence in making a discrimination response (e.g. rightward vs. leftward) increases away from the major diagonal (Figure \ref{fig:fig4}B), as the model becomes sure that the sample was generated by either a leftward or rightward tilted stimulus. As in \cite{King:2014}, these changes in discrimination confidence may still occur in the absence of reporting ``seen".

I next simulate a proxy for prediction error at each layer in the model -- in other words, how much belief change was induced by the sensory sample. I use the Kullback-Leibler (K-L) divergence as a compact summary of how far the posterior probability distribution at each level in the network differs from the prior. Flat priors were used for both the $A$ and $W$ levels. The K-L divergence is a measure of Bayesian surprise at each level in the network, which under predictive coding accounts is linked to neural activation at each level in a hierarchical network \cite{Friston:2005, Summerfield:2014}. For instance, if the model starts out with a strong prior that it will see gratings of either orientation, but a grating is omitted, this constitutes a large prediction error (an unexpected absence). Thus computing K-L divergence for different levels of the network provides a rough proxy for the amount of ``activation" we would expect as a function of different types of decision.

\begin{figure}
  \centering
  \includegraphics[width=160mm]{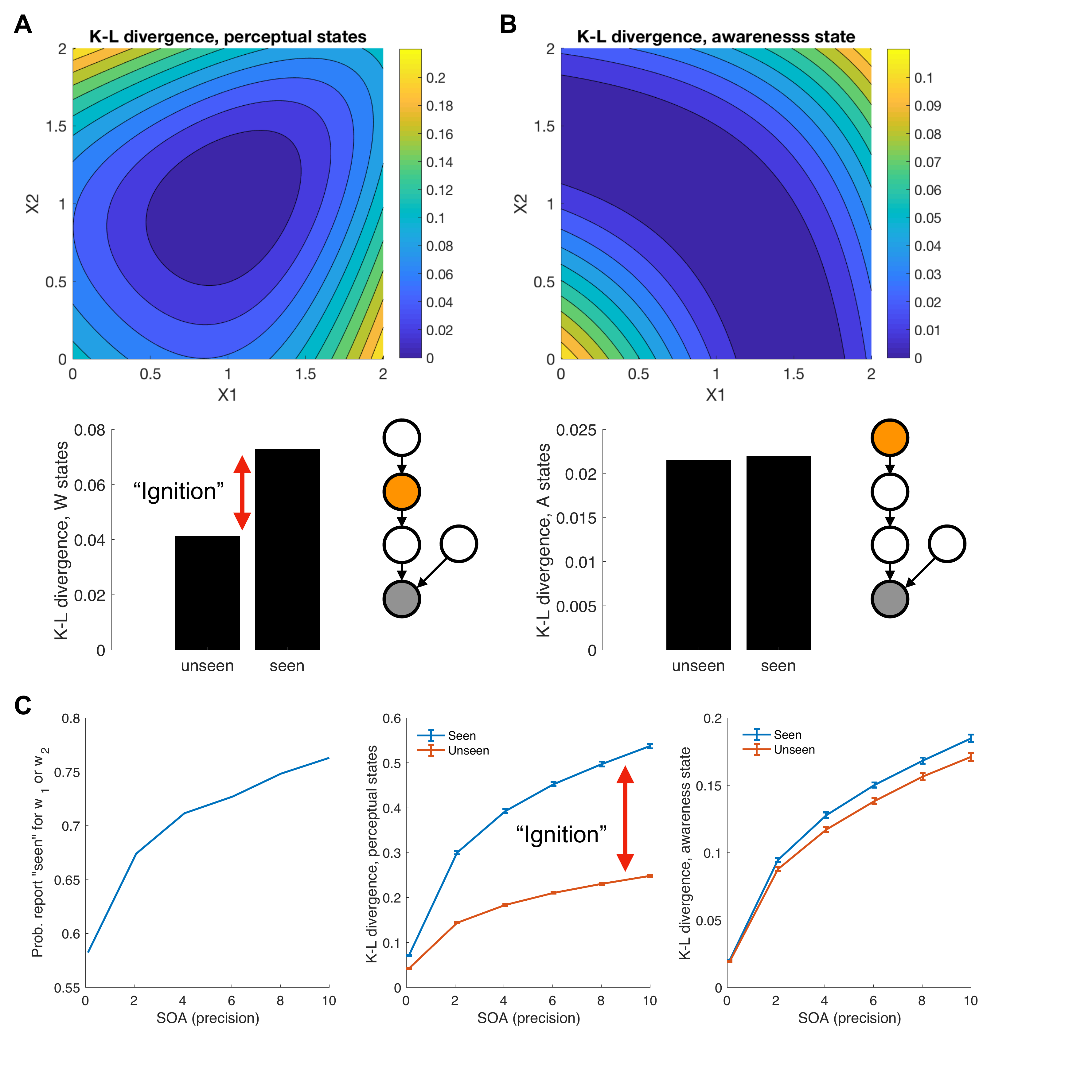}
  \caption{Figure 5. A, B) Kullback-Leibler (K-L) divergence for A) perceptual states $W$ and B) awareness state $A$ as a function of sensory input $X$. K-L divergence quantifies the change from prior to posterior after seeing the stimulus $X$, and provides a metric for the magnitude of belief update at different levels of the network. The lower panels show the averaged K-L divergence for both $W$ and $A$ as a function of whether the model reports presence ($P(A=a_1|X=x) > 0.5$) or absence. The network nodes correspond to those in Figure \ref{fig:fig2}A and the orange node indicates the node for which the K-L divergence is calculated. C) Behaviour of the network in a simulated masking experiment at various levels of stimulus-onset asynchrony (SOA, modeled as increasing sensory precision) in which sensory evidence was sampled from the three stimulus classes shown in Figure \ref{fig:fig3}B. The lefthand panel shows that the model is more likely to report ``seen" as SOA increases. The middle panel shows the K-L divergence at the level of perceptual states $W$ as a function of whether the model reports presence ($P(A=a_1|X=x) > 0.5$) or absence. The expected K-L divergence is asymmetric, with a bigger average belief update following ``seen" decisions (a computational correlate of global ignition). The righthand panel shows the average KL divergence of awareness state $A$ as a function of whether the model reports presence ($P(A=a_1|X=x) > 0.5$) or absence. At this level the expected K-L divergence is relatively symmetric for ``seen" and ``unseen" decisions.}
  \label{fig:fig5}
\end{figure}

At the level of perceptual states $W$, there is substantial asymmetry in the K-L divergence expected when the model says ``seen" vs. ``unseen" (Figure \ref{fig:fig5}A). This is due to the large belief updates invoked in the perceptual layer $W$ by samples that deviate from the origin. In contrast, when we compute K-L divergence for the awareness state (Figure \ref{fig:fig5}B), the level of prediction error is symmetric across seen and unseen decisions. This is because at this level there is minimal asymmetry between inference on presence and absence. When simulating these belief updates over a range of precisions to mimic increasing stimulus-onset asynchrony in a typical backward-masking experiment, we see that the asymmetry in K-L divergence of the $W$ states increases with SOA, producing an ignition-like pattern when the stimulus is ``seen" (Figure \ref{fig:fig5}C).

\section{Empirical predictions}

The model is currently situated at a computational level and remains agnostic about temporal dynamics and neural implementation\footnote{For recent work translating probabilistic graphical models into models of neuronal message passing see \cite{George:2017, Friston:2017}.}. Here I instead focus on coarser-scale predictions about the neural correlates of awareness reports in typical consciousness experiments.

First, as hinted above, an asymmetric state space for presence and absence suggests there will be greater summed prediction error in the entire network on presence decisions (as summarised by K-L divergence at each node of $W$). This may be a computational correlate of the global ignition responses often found to track awareness reports \cite{Del-Cul:2007, Dehaene:2011}.

Second, the model predicts that awareness reports (but not discrimination performance, which relies on lower-order inference on $W$) will depend on higher-order states. These may be instantiated in neural populations in prefrontal and parietal cortex \cite{Lau:2011}. Thus it may be possible to silence or otherwise inactivate the neural substrates of an awareness state without affecting performance -- a type of blindsight \cite{Del-Cul:2009, Weiskrantz:1999}. However, to the extent that this network is flexible in its functional contribution to higher cognition, showing both ``multiple demand" characteristics \cite{Duncan:2010} and mixed selectivity \cite{Mante:2013}, we should also not be surprised by null results, given that single lesions may belie redundancy in its contribution to awareness \cite{Michel:}.

\begin{figure}
  \centering
  \includegraphics[width=160mm]{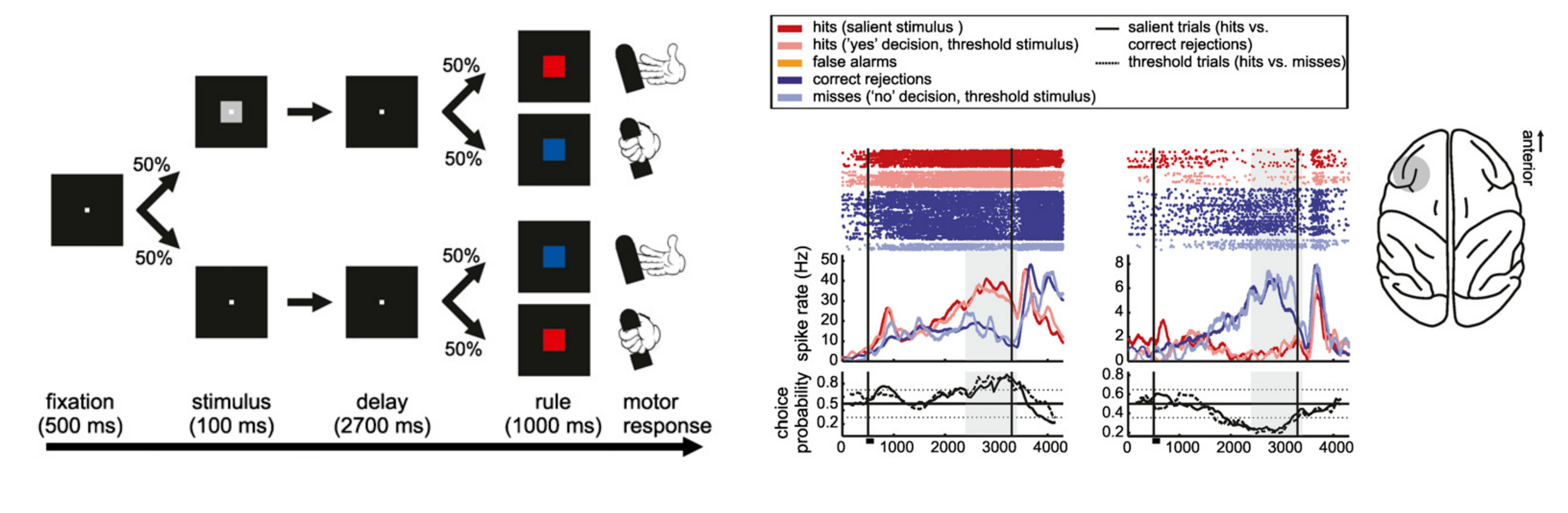}
  \caption{Figure 6. Experimental paradigm and sample results reproduced from Merten and Nieder \cite{Merten:2012}. The left-hand panel shows the experimental paradigm used to study decisions about stimulus absence and presence after controlling for sensory and motor features of the response. The monkeys initiated each trial by grasping a lever and fixating a central fixation target. A low-contrast stimulus was then displayed for 100ms (on 50\% of trials) or a blank screen was maintained (on the other 50\% of trials). After a short delay, the response mappings for that trial were revealed (on some trials a present decision would require a lever release, whereas on other trials the same decision would require a lever hold). The right-hand panel shows that firing rates of neural populations in LPFC tracked abstract decisions about presence or absence before the motor mapping was known, and did so independently of stimulus properties (similar activations were seen for hits and false alarms, and for misses and correct rejections).}
  \label{fig:fig6}
\end{figure}

Third, for the uppermost awareness state, we expect symmetry -- decisions in favour of both presence and absence will lead to belief updates of similar magnitude. There has been limited focus on examining decisions about stimulus \emph{absence} (as these decisions are often used as a baseline or control condition in studies of perceptual awareness). However, existing data are compatible with symmetric encoding of presence and absence at the upper level of the visual hierarchy, in primate lateral prefrontal cortex (LPFC; \cite{Panagiotaropoulos:2012}). Merten and Nieder \cite{Merten:2012} trained monkeys to report the presence or absence of a variety of low-contrast shapes presented near to visual threshold. Neural activity tracking the decision (present or absent) was decorrelated from that involved in planning a motor response by use of a post-stimulus cue that varied from trial to trial (Figure \ref{fig:fig6}). Distinct neural populations tracked the decision to report ``seen" vs. ``unseen". Importantly the magnitude of activation of these populations was similar in timing and strength, suggesting a symmetric encoding of awareness in LPFC. Using fMRI, Christensen et al.\ also observed symmetric activation for judgments of presence and absence (compared to an intermediate visibility rating) in anterior prefrontal cortex, whereas a global ignition response was seen for presence (compared to absence) in a widespread frontoparietal/striatal network \cite{Christensen:2006}.

More broadly, the current framework suggests that focusing on inference about absence will be particularly fruitful for understanding the neural and computational basis of awareness reports \cite{Farennikova:2013, Kanai:2010, Martin:2013, Merten:2012, Merten:2013}.

\section{Role of precision estimation in resolving awareness states}

The current simulations assume that the noise of sensory input is fixed, but in reality this parameter would also need to be estimated. A range of disambiguating cues are likely to prove important in such an estimation scheme and thereby affect an inference on awareness.  For instance, beliefs about the state of attention or other properties of the sensory system provide global, low-dimensional cues as to the state of awareness \cite{Lau:2008, Graziano:2013}. Computationally these cues may be implemented as beliefs about precision (priors on $\Sigma$ in the model in Figure \ref{fig:fig2}A), where precision refers to the inverse of the noise (variability) we expect from a particular sensory channel.

Another source of information about sensory precision is proprioceptive and interoceptive input about bodily states. Consider the following thought experiment in which we set up two conditions in a dark room, one in which the subject has their eyes open and one in which they have their eyes closed. Now imagine that we have arranged for neural activity in early visual areas to be identical in the two cases (the $X$'s are the same), and that in both cases the subject is told (for instance via an auditory cue) that there might have been a faint flash of light. Despite the visual activity being identical, the subject can be sure that they didn't see anything when their eyes were closed compared to when they were open. In other words, prioprioception provides disambiguating information as to the current state of awareness -- when the eyes are open we expect to have higher precision input than when the eyes are closed.

One straightforward way of introducing this relationship is to allow precision itself to depend on awareness (a connection between $A$ and $\Sigma$ in Figure \ref{fig:fig2}A. Such a modification implies that an awareness state may be two-dimensional, encoding the distinction between whether something has the \emph{potential} to be seen (high vs. low expected precision or eyes open vs. eyes closed) as well as whether something \emph{is} seen (present vs. absent; \cite{Metzinger:2014, Limanowski:2018}). This aspect of the HOSS model is also in keeping with Graziano's attention schema model of consciousness, in which awareness is equated to a model of attention \cite{Graziano:2013}. However, in contrast to the attention schema approach, in HOSS a model of attention provides a critical \emph{input} into resolving ambiguity about whether we are aware or not (by affecting beliefs about precision), rather than determining awareness itself.

\section{Relationship to other theories of consciousness}

The goal of the higher-order state-space (HOSS) approach outlined here is modest - to delineate computations supporting metacognitive \emph{reports} about awareness. This is a useful place to start given that report (or the potential for report) is the jumping-off point for a scientific study of consciousness.

A stronger reading of the model is that conscious awareness and metacognitive reports depend on shared mechanisms in the human brain \cite{Shaver:2008, Brown:2019}. This stronger version shares similarities with higher-order theories of consciousness, particularly Lau's proposal that consciousness involves ``signal detection on the mind" \cite{Lau:2008, Hohwy:2015}. Notably, a process of hierarchical inference may take place via passive message-passing without any strategic, cognitive access to this information e.g. in working memory \cite{Carruthers:2017}, making it compatible with higher-order representational accounts of phenomenal consciousness \cite{Brown:2015}.

However, while inference on higher-order states is, on this view, necessary for awareness, it may not be sufficient. In HOSS, the higher-order awareness state is simple and low-dimensional. Lower-order states clearly must make a contribution to perceptual experience under this arrangement -- a variant of the ``joint determination" view advocated by Lau and Brown \cite{Lau:2019}. However it seems an empirical question as to the relative granularity of higher-order and first-order representations in terms of their contribution to conscious experience, and a range of intermediate views are plausible. The more important point is that the state space is factorised to allow two separate causes of the sensory data -- what it is, and whether I have seen it. In other words, becoming aware of a red, tilted object may depend on learning an abstract, factorised state of presence/absence that is not bound up with the states of being red or tilted. This is likely to be computationally demanding in a rich, multidimensional state space (as it requires marginalising over $W$), and may be rare in cognitive systems.

HOSS also provides a new perspective on global workspace (GWS) architectures. GWS proposes that consciousness occurs when information is ``globally broadcast" throughout the brain. As a result of global broadcast, cognitive and linguistic machinery have access to information about a particular stimulus or subpersonal mental state \cite{Dehaene:1998}. HOSS retains the ``global" aspect of GWS, in that an awareness state is hierarchically higher with respect to the range of possible perceptual states, and therefore has a wide conceptual purview. However, HOSS recasts ignition-like activations as asymmetric inference about stimulus presence rather than a consequence of stimulus content being ``broadcast". In any case, it is arguable whether such global broadcast is able to directly account for how a system claims to be conscious of a stimulus without positing additional machinery. Global access to the workspace would allow the system to say ``there is an \emph{X}", but not endow it with the capacity to report awareness of \emph{X}. This point is made concisely by Graziano \cite{Graziano:2016}:

\begin{quotation}
Consider asking `Are you aware of the apple?' The search engine searches the internal model and finds no answer. It finds information about an apple, but no information about what `awareness' is, or whether it has any of it... It cannot answer the question. It does not compute in this domain.
\end{quotation}

In other words, it is difficult to see how such a system can be actively aware of the absence of stimulation when global broadcast is constitutive of awareness. The state space approach outlined here is designed explicitly to compute in this domain, and therefore does not suffer from the same problem. Another critical difference between GWS and HOSS is that HOSS predicts prefrontal involvement for active decisions about stimulus \emph{absence}, whereas GWS predicts that PFC remains quiescent on such trials due to a failure of the stimulus to gain access to the workspace.

Finally, to the extent that abstract awareness states need to be learnt or constructed, creating this level may require a protracted period of development. Such development would begin with creating a perceptual generative model ($W$) before a more general property (awareness) could be abstracted from these perceptual states. This is consistent with Cleeremans' ``radical plasticity thesis" in which consciousness is underpinned by learning abstract representations of both ourselves and the world \cite{Cleeremans:2011}.

\section{Research questions}

I close with questions for future research motivated by the current computational sketch:

\begin{enumerate}
\item How are awareness states represented in neural activity? Are presence and absence encoded symmetrically?
\item Is a (neural) representation of awareness factorised with respect to other aspects of perceptual content?
\item How are awareness states learned?
\end{enumerate}

\section*{Acknowledgements}

I am grateful to the Metacognition and Theoretical Neurobiology groups at the Wellcome Centre for Human Neuroimaging and members of the University of London Institute of Philosophy for helpful discussions. I thank Matan Mazor, Oliver Hulme, Chris Frith, Nicholas Shea, Karl Friston and two anonymous reviewers for comments on previous drafts of this manuscript. This work was supported by a Wellcome / Royal Society Sir Henry Dale Fellowship (206648/Z/17/Z).

\section*{Data availability}

Model and simulation code can be accessed at \url{https://github.com/smfleming/HOSS}.

\bibliographystyle{unsrt}
\bibliography{statespacebib2}

\end{document}